\documentclass{bio}
\usepackage[T1]{fontenc}
\usepackage[utf8]{inputenc}
\usepackage{url}
\usepackage{float}
\makeatletter
\def\ps@plain{%
  \def\@oddhead{}%
  \def\@evenhead{}%
  \def\@oddfoot{}%
  \def\@evenfoot{}%
}
\def\ps@headings{%
  \def\@oddhead{}%
  \def\@evenhead{}%
  \def\@oddfoot{}%
  \def\@evenfoot{}%
}
\makeatother

\begin{document}
\vspace*{-4cm}

\title{A Spatio-Temporal Self-Propagating Log-Gaussian Cox-Hawkes Process for Star Formation Modelling}
\author{
Qihan Zou$^{1,2}$ \\[4pt]
$^{1}$\textit{University of Melbourne}\\
$^{2}$\textit{RMIT University}\\[2pt]
\texttt{qihanzou@alumni.unimelb.edu.au}
}

\maketitle
\thispagestyle{empty}
\begin{center}
\textbf{Abstract}
\end{center}
\noindent
Stochastic self-propagating star formation models describe galactic structure through local triggering and feedback but are commonly formulated using discrete spatial cells and time steps. We extend the spatio-temporal log-Gaussian Cox--Hawkes framework to obtain the Spatio-Temporal Self-Propagating Log-Gaussian Cox--Hawkes Process, a continuous point process model for the locations and times of star-forming events. The model combines spontaneous formation, correlated environmental effects, outwardly propagating excitation, local inhibition, differential rotation, and saturation. An observation layer transforms the conditional event intensity into idealised maps of instantaneous star-forming arm emissivity and recent young stellar surface brightness. For the selected parameter setting, the displayed realisation exhibits transient flocculent spiral-like patterns without any deterministic spiral geometry being imposed. The Monte Carlo experiment finds similar event production with and without rotation, whereas the stationary-front case produces fewer events. The displayed maps further suggest that differential rotation contributes to the winding and spatial arrangement of activity. The proposed framework provides a basis for future statistical inference from spatially and temporally resolved observations of star formation.


\section{Introduction}
Classical density wave theories describe spiral arms through large scale wave-like organisation of the disk \citep{lin1979density,lin1987spiral}, whereas an alternative view treats galactic morphology as the collective outcome of local and short-lived star-formation activity. An early propagating star-formation model was developed by \citet{mueller1976propagating}, with a stochastic formulation subsequently introduced by \citet{gerola1978stochastic}. In SSPSF models, star formation can occur spontaneously or be triggered near recently active regions, while feedback temporarily suppresses repeated formation at the same location. Differential rotation then transports and shears the resulting activity, allowing local stochastic interactions to generate large-scale, irregular spiral patterns without prescribing spiral arms in advance. Later variants incorporated radial changes in the star-formation mechanism, three-dimensional disk structure, and anisotropic propagation \citep{comins1981profiles,statler1983stochastic,jungwiert1994stochastic}. Point processes describe the locations and times of random events through a conditional intensity. The Cox process of \citet{cox1955some} is conditionally Poisson given a stochastic intensity, providing a representation of clustering associated with unobserved environmental effects. \citet{moller1998log} formulated the log-Gaussian Cox process by assigning a Gaussian random field to the log-intensity, and \citet{brix2001spatiotemporal} developed spatio-temporal prediction for this model class. A complementary approach was introduced by \citet{hawkes1971spectra}, whose self-exciting process allows previous events to increase future conditional intensity. Its immigration--birth representation was established by \citet{hawkes1974cluster}, while \citet{ogata1998space} developed an influential spatio-temporal formulation for earthquake occurrences. These two modelling traditions provide complementary descriptions of latent environmental variation and event-driven dependence. A direct combination is the Cox--Hawkes model of \citet{miscouridou2022cox}, which uses a log-Gaussian Cox background together with additive Hawkes triggering.

Within a log-intensity formulation, \citet{zou2025spatio} connected the SSPSF mechanism to spatio-temporal point processes and introduced signed history dependence. The sign of a single coefficient $\alpha$ determines whether this dependence is excitatory or inhibitory, so outward excitation and local recovery cannot operate simultaneously. The temporal contribution is absent until the event reaches age $\rho$ and then begins abruptly before decaying monotonically. Thus, $\rho$ acts as a feedback-onset delay and does not define recovery at the affected location. Spatially, each contribution is represented by a Gaussian kernel centred on the event's fixed birth location, without an expanding front or differential rotation transport. To incorporate these mechanisms and explore their implications for galactic morphology, we develop the Spatio-Temporal Self-Propagating Log-Gaussian Cox--Hawkes Process (STSP--LGCHP) as an extension of this continuous framework. The model separates history dependence into an outwardly propagating excitation field and a local inhibitory field for region recovery, allowing the two effects to operate simultaneously. A shifted Erlang-2 temporal kernel provides a continuous feedback onset, differential rotation transports and shears the excitation fronts, and saturation limits the combined positive feedback from overlapping fronts. The simulation study investigates whether this construction can produce transient flocculent spiral-like star-formation patterns without imposing deterministic spiral geometry and compares the event counts and spatial patterns obtained under alternative front and rotation specifications.

The remainder of this paper is organised as follows. Section 2 reviews the underlying log-Gaussian Cox--Hawkes framework and defines the STSP--LGCHP. Section 3 presents the simulation study, and Section 4 discusses its interpretation, limitations, and directions for future work.

\section{Spatio-Temporal Log-Gaussian Cox-Hawkes Processes (LGCHP)}
\subsection{The Spatio-Temporal Log-Gaussian Cox-Hawkes Process Framework}\label{subsec:original-lgchp}
Let $\mathcal{S}\subset\mathbb{R}^{2}$ be a bounded spatial domain, let $[0,T]$ be the observation interval, and let $\mathcal{H}_t=\{(s_i,t_i):t_i<t\}$ denote the event history before time $t$. The log-intensity LGCHP formulation of \citet{zou2025spatio} has conditional intensity $$\lambda(s,t\mid\mathcal{H}_t,Y)=\exp\left\{\mu(s,t)+Y(s,t)+D(s,t\mid\mathcal{H}_t)\right\},$$ where $\mu(s,t)$ is a deterministic log-background intensity, $Y(s,t)$ is a latent Gaussian field, and $D(s,t\mid\mathcal{H}_t)$ represents dependence on previous events. The latent field is specified as $Y\sim\operatorname{GP}(0,C_Y)$ with covariance $$C_Y\left\{(s_1,t_1),(s_2,t_2)\right\}=\sigma_Y^2\exp\left\{-\frac{\lVert s_2-s_1\rVert}{\phi_Y}-\theta_Y|t_2-t_1|\right\}.$$ The history-dependent component is $$D(s,t\mid\mathcal{H}_t)=\alpha\sum_{i:t_i<t}\exp\left\{-\beta(t-t_i)\right\}\varphi_{\Sigma}(s-s_i),$$ where $$\varphi_{\Sigma}(h)=\frac{1}{2\pi|\Sigma|^{1/2}}\exp\left\{-\frac{1}{2}h^\top\Sigma^{-1}h\right\}.$$ Here, $\alpha\in\mathbb{R}$ determines the sign and strength of the history effect, with $\alpha>0$ corresponding to excitation and $\alpha<0$ to inhibition, while $\beta>0$ controls temporal decay and $\Sigma$ controls the spatial scale and shape of the interaction. \citet{zou2025spatio} also introduced the delayed form \(D_{\rho}(s,t\mid\mathcal{H}_t)=\alpha\sum_{i:t_i<t-\rho}\exp\left\{-\beta(t-t_i)\right\}\varphi_{\Sigma}(s-s_i),\) where $\rho>0$ is the delay before an event affects the intensity.

\subsection{The Spatio-Temporal Self-Propagating Log-Gaussian Cox--Hawkes Process}\label{subsec:stsp-lgchp}
Let $s\in\mathcal{S}\subset\mathbb{R}^2$ denote a spatial location in a bounded galactic disc, let $t\in[0,T]$ denote time, and let $\mathcal{H}_t=\{(s_i,t_i):t_i<t\}$ denote the history of star formation events immediately before time $t$. Each event $(s_i,t_i)$ represents the birth of a star-forming region at location $s_i$ and time $t_i$, rather than the formation of an individual star. Building on the spatio-temporal log-Gaussian Cox--Hawkes process introduced in Section~\ref{subsec:original-lgchp}, we propose the Spatio-Temporal Self-Propagating Log-Gaussian Cox--Hawkes Process (STSP-LGCHP). Its conditional intensity is $$\lambda(s,t\mid\mathcal{H}_t,Y)=\kappa g_0(s)\exp\left\{Y(s,t)-\frac{\sigma_Y^2}{2}+D_+(s,t)-D_-(s,t)\right\},$$ where $\kappa>0$ is the baseline parameter, $g_0(s)$ is a deterministic spatial background, $Y(s,t)$ is a latent Gaussian field representing unobserved environmental variation, $D_+(s,t)$ is an outwardly propagating excitation field, and $D_-(s,t)$ is a local inhibitory field governing regional recovery. The resulting conditional intensity has units of events per unit area per unit time. We specify the deterministic background as $$g_0(s)=\frac{1}{2\pi|\Gamma_0|^{1/2}}\exp\left\{-\frac{1}{2}(s-m_0)^\top\Gamma_0^{-1}(s-m_0)\right\},$$ where $m_0\in\mathbb{R}^2$ is the centre of the galactic disc and the symmetric positive-definite matrix $\Gamma_0\in\mathbb{R}^{2\times2}$ controls the spatial scale, orientation and anisotropy of the background. The Gaussian density $g_0$ is defined on $\mathbb{R}^2$ and evaluated on $\mathcal{S}$ without being renormalised over the bounded observation region. Consequently, $\kappa$ is a baseline parameter with units of events per unit time and is not equal to the total background event rate within $\mathcal{S}$. The latent environmental field is specified as $$Y\sim\operatorname{GP}(0,C_Y),$$ with covariance function $$C_Y\left\{(s,t),(s',t')\right\}=\sigma_Y^2\exp\left\{-\frac{\lVert s-s'\rVert}{\phi_Y}-\theta_Y|t-t'|\right\}.$$ The parameter $\sigma_Y\geq0$ is the marginal standard deviation of the latent field on the log-intensity scale, $\phi_Y>0$ is its spatial correlation scale, and $\theta_Y>0$ is the temporal decorrelation rate of the latent field. The correction $-\sigma_Y^2/2$ ensures that $$\mathbb{E}\left[\exp\left\{Y(s,t)-\frac{\sigma_Y^2}{2}\right\}\right]=1.$$ The term $\kappa g_0(s)$ determines the baseline rate of spontaneous events, while $Y(s,t)$ represents correlated environmental effects across space and time.

To represent differential rotation, define the galactocentric radius of $s$ by $$r(s)=\lVert s-m_0\rVert,$$ and let the angular speed be $$\omega\{r(s)\}=\frac{\omega_c}{\sqrt{1+\{r(s)/r_c\}^2}},$$ where $\omega_c\geq0$ is the central angular-speed scale and $r_c>0$ is the characteristic radius over which the rotation curve changes. The associated flow map is $$\Phi_u(s)=m_0+R_{\omega\{r(s)\}u}(s-m_0),$$ where $u$ is a time displacement and $$R_\vartheta=\begin{pmatrix}\cos\vartheta&-\sin\vartheta\\\sin\vartheta&\cos\vartheta\end{pmatrix}$$ denotes planar rotation through angle $\vartheta$. This rotation law produces approximately rigid rotation near the centre and a decreasing angular speed at larger radii. For a previous event $(s_i,t_i)$, let $$u_i=t-t_i>0$$ denote its age at time $t$. We define its rotation-adjusted displacement from a candidate location $s$ by $$h_i(s,t)=\Phi_{-u_i}(s)-s_i,$$ and the corresponding rotation-adjusted distance by $$d_i(s,t)=\lVert h_i(s,t)\rVert.$$ The quantity $d_i(s,t)$ is obtained by mapping the current candidate location backwards through the differential-rotation flow before measuring its distance from the event's birth location. Consequently, the history interaction is transported and sheared by the galactic flow rather than remaining fixed in the original Euclidean coordinates.

The temporal amplitude of the positive feedback is described by the peak-normalised shifted Erlang-2 kernel $$q_{\rho,\beta}(u)=\mathrm{e}\,\beta(u-\rho)\exp\left\{-\beta(u-\rho)\right\}\mathbf{1}\{u>\rho\},$$ where $\mathrm{e}=\exp(1)$, $\mathbf{1}\{\cdot\}$ is the indicator function, $\rho\geq0$ is the feedback-onset delay, and $\beta>0$ controls the subsequent rise and decay of the feedback effect. The kernel begins continuously from zero at $u=\rho$ and reaches its unit maximum at $$u=\rho+\frac{1}{\beta}.$$ The case $\rho=0$ corresponds to a feedback effect that begins immediately but still rises continuously from zero. The temporal kernel $q_{\rho,\beta}$ controls the amplitude of the feedback, whereas the propagation speed introduced below controls the movement of the feedback front. These are distinct components of the model. After the onset time, the positive effect of event $i$ is concentrated around an expanding annular front, $$F_i(s,t)=\exp\left[-\frac{\left\{d_i(s,t)-v_f(u_i-\rho)_+\right\}^2}{2w_f^2}\right],$$ where $(x)_+=\max(x,0)$, $v_f\geq0$ is the propagation speed, and $w_f>0$ is the width of the front. At event age $u_i$, the front is centred at the propagation radius $$r_i^{\mathrm{front}}(t)=v_f(u_i-\rho)_+.$$ When the shifted Erlang-2 temporal envelope reaches its maximum, the corresponding front radius is $r_i^{\mathrm{front}}=\frac{v_f}{\beta}.$ The front kernel is peak-normalised rather than spatially normalised, so $F_i(s,t)=1$ whenever $d_i(s,t)=v_f(u_i-\rho)_+$. The unsaturated positive history field is $$D_{+,\mathrm{raw}}(s,t)=\alpha_+\sum_{i:t_i<t}q_{\rho,\beta}(u_i)F_i(s,t),$$ where $\alpha_+\geq0$ controls the peak contribution of an isolated event to the unsaturated log-intensity field. Since several propagating fronts may overlap, directly exponentiating $D_{+,\mathrm{raw}}$ could produce extremely large conditional intensities. We therefore apply the aggregate saturation transformation $$D_+(s,t)=c_{\mathrm{s}}\left[1-\exp\left\{-\frac{D_{+,\mathrm{raw}}(s,t)}{c_{\mathrm{s}}}\right\}\right],$$ where $c_{\mathrm{s}}>0$ is the positive saturation level. The bounds $0\leq D_+(s,t)<c_{\mathrm{s}}$ and $D_-(s,t)\geq0$ imply $$\lambda(s,t\mid\mathcal{H}_t,Y)\leq\kappa g_0(s)\exp\left\{Y(s,t)-\frac{\sigma_Y^2}{2}+c_{\mathrm{s}}\right\}.$$ Conditional on a bounded realisation of $Y$ over the bounded observation window, the right-hand side is integrable, and the conditional process is therefore non-explosive on $\mathcal{S}\times[0,T]$. The parameter $\alpha_+$ is the peak unsaturated contribution of a single event. Saturation bounds the combined positive history effect by $c_{\mathrm{s}}$ and is weak when $D_{+,\mathrm{raw}}(s,t)/c_{\mathrm{s}}$ is small.

The local inhibitory field is defined as $$D_-(s,t)=\alpha_-\sum_{i:t_i<t}\exp\left(-\frac{u_i}{\tau_R}\right)\exp\left\{-\frac{d_i(s,t)^2}{2\ell_R^2}\right\},$$ where $\alpha_-\geq0$ controls the initial strength of the local suppression, $\tau_R>0$ is its temporal decay scale, and $\ell_R>0$ is its spatial range. Since $D_-$ enters the log-intensity with a negative sign, a newly generated event immediately reduces the subsequent formation rate near the corresponding material location. This suppression weakens over time, while the delayed positive effect develops and propagates outwards. The combined mechanism therefore discourages repeated formation at a recently active location and promotes subsequent formation near the expanding front. The complete model parameter set is $\Theta=\left(\kappa,m_0,\Gamma_0,\sigma_Y,\phi_Y,\theta_Y,\alpha_+,\rho,\beta,v_f,w_f,c_{\mathrm{s}},\alpha_-,\tau_R,\ell_R,\omega_c,r_c\right).$ The parameters $\rho$ and $\tau_R$ have units of time; $\beta$, $\theta_Y$, and $\omega_c$ have units of inverse time; $v_f$ has units of distance per unit time, $m_0$, $\phi_Y$, $w_f$, $\ell_R$, and $r_c$ have units of distance and $\Gamma_0$ has units of squared distance. The parameters $\sigma_Y$, $\alpha_+$, $\alpha_-$, and $c_{\mathrm{s}}$ are dimensionless log-intensity contributions. Each new event generates a delayed excitation front and a local inhibitory region, allowing activity to continue across event generations. No spiral template is included in the Gaussian background or elsewhere in the model. Spiral-like features can instead emerge through differential shearing, local inhibition, and repeated event generation and are expected to be transient and flocculent rather than prescribed grand-design arms.

\section{Simulation Study}\label{sec:simulation}
We study the behaviour of the STSP--LGCHP through simulation using the event dynamics specified in Section~\ref{subsec:stsp-lgchp}. We consider the circular spatial domain $\mathcal{S}=\{s\in\mathbb{R}^2:\lVert s-m_0\rVert\leq R\}$, where $m_0=(0.5,0.5)^\top$ and $R=0.48$, over the observation period $[0,T]$ with $T=1000$. The latent Gaussian field is approximated as constant within each cell of a regular space--time grid. We use a $100\times100$ spatial grid and $1000$ temporal cells, giving $\Delta t_Y=1$. Field values at the spatial cell centres are generated sequentially using the spatial covariance specified by $C_Y$ and temporal correlation $\exp(-\theta_Y\Delta t_Y)$. Conditional on the generated field, continuous event times and locations are simulated by thinning. Within temporal cell $k$, the pointwise proposal bound is $$\overline{\lambda}_k(s)=\kappa g_0(s)\exp\left\{\max_jY_{kj}-\frac{\sigma_Y^2}{2}+c_{\mathrm{s}}\right\},$$ where $Y_{kj}$ is the field value in spatial cell $j$. Since $g_0$ integrates to one over $\mathbb{R}^2$, the corresponding integrated proposal rate is $$\overline{\Lambda}_k=\int_{\mathbb{R}^2}\overline{\lambda}_k(s)\,ds=\kappa\exp\left\{\max_j Y_{kj}-\frac{\sigma_Y^2}{2}+c_{\mathrm{s}}\right\}.$$ Candidate times are generated using $\overline{\Lambda}_k$, and candidate locations are drawn from $g_0$. Locations outside $\mathcal{S}$ are discarded, while a remaining candidate $(s,t)$ is accepted with probability $\frac{\lambda(s,t\mid\mathcal{H}_{t^-},Y)}{\overline{\lambda}_k(s)}.$ The event history is updated after each accepted event, and simulation continues until time $T$.

The remaining baseline parameters are $\kappa=0.04$ and $\Gamma_0=0.08I_2$ for the deterministic background, and $\sigma_Y=0.20$, $\phi_Y=0.12$, and $\theta_Y=0.04$ for the latent Gaussian field. The positive history parameters are $\alpha_+=3.6$, $\rho=4$, and $\beta=0.05$. The propagation speed and front width are $v_f=0.004$ and $w_f=0.020$, respectively, and the saturation level is $c_{\mathrm{s}}=7$. The local inhibitory field uses $\alpha_-=2.5$, $\tau_R=100$, and $\ell_R=0.020$, while differential rotation uses $\omega_c=0.045$ and $r_c=0.10$. The observation layer uses $A=60$, $\tau_L=20$, $h_b=0.012$, and $\Delta a=5$. All spatial and temporal coordinates are dimensionless, and no conversion to physical units is assumed. Six pre-observation events are assigned times $-80,-65,-50,-35,-20,$ and $-5$, with locations independently sampled from $g_0$ conditional on lying in $\mathcal S$. They form the initial history at $t=0$, are not counted among events generated during $[0,T]$, and contain no deterministic spiral arrangement. A representative realisation is evaluated at $t=400$, $600$, $800$, and $1000$.

Let $\lambda_t(s)=\lambda(s,t\mid\mathcal{H}_{t^-},Y)$ denote the conditional star formation event intensity defined by the stochastic model. For a small spatial region $ds$ and a short interval $dt$, the conditional probability of an event is approximately $\lambda_t(s)\,ds\,dt$. For an evaluation location $x\in\mathcal S$ and bandwidth $h>0$, define the unit-integral Gaussian kernel $$K_h(z)=\frac{1}{2\pi h^2}\exp\left(-\frac{\lVert z\rVert^2}{2h^2}\right).$$ The parameter $h_b>0$ is the observation bandwidth. The relative instantaneous star forming arm emissivity is defined as $$E_{\mathrm{inst}}(x,t)=\int_{\mathcal{S}}K_{h_b}(x-s)\lambda_t(s)\,ds.$$ This idealised map smooths the event-rate surface to represent the finite spatial extent of star forming regions and observational resolution, without affecting the event dynamics. To represent emission from recently formed young stellar populations, we introduce the age response function $$\ell(a)=\exp(-a/\tau_L)\mathbf{1}\{0\leq a\leq A\},$$ where $A$ is the maximum retained age and $\tau_L$ controls luminosity decay. The relative young stellar surface brightness is defined from the conditional intensity as $$B_\lambda(x,t)=\int_0^{\min(A,t)}\ell(a)\int_{\mathcal{S}}K_{h_b}\left\{x-\Phi_a(s)\right\}\lambda_{t-a}(s)\,ds\,da.$$ This field represents idealised expected emission from recent star formation and is not a realised event catalogue, stellar mass density, or gas density. Using midpoint quadrature with $a_j=(j+1/2)\Delta a$ and $J=A/\Delta a$, we obtain $$\widehat B_\lambda(x,t)=\Delta a\sum_{\substack{0\leq j<J\\a_j<\min(A,t)}}\exp(-a_j/\tau_L)\int_{\mathcal{S}}K_{h_b}\left\{x-\Phi_{a_j}(s)\right\}\lambda_{t-a_j}(s)\,ds.$$ 

For each plotted quantity, a common colour scale is used across all cases and evaluation times, with the upper display limit set to the pooled $99.5$th percentile. We compare the baseline process with two kinematic modifications. The no-rotation case sets $\omega_c=0$ in both the history transport and the observation-layer advection. In the stationary-front case, the expanding radius $v_f(u_i-\rho)_+$ is replaced by the fixed radius $r_\ast=v_f/\beta=0.08$, corresponding to its baseline value at the peak of the temporal envelope. All other parameters, the latent field, and the initial history are held fixed across the three cases. Because $F_i$ is peak-normalised, fixing its radius also changes the excitation integrated over space. The comparison therefore concerns the complete expanding-front and stationary-front specifications. A no-recovery case is outside the scope of the present study. The maps below show one paired realisation and provide a qualitative comparison of morphology. The Monte Carlo analysis later summarises event counts across repeated simulations.

\begin{figure}[ht]
    \centering
    \includegraphics[width=1\linewidth]{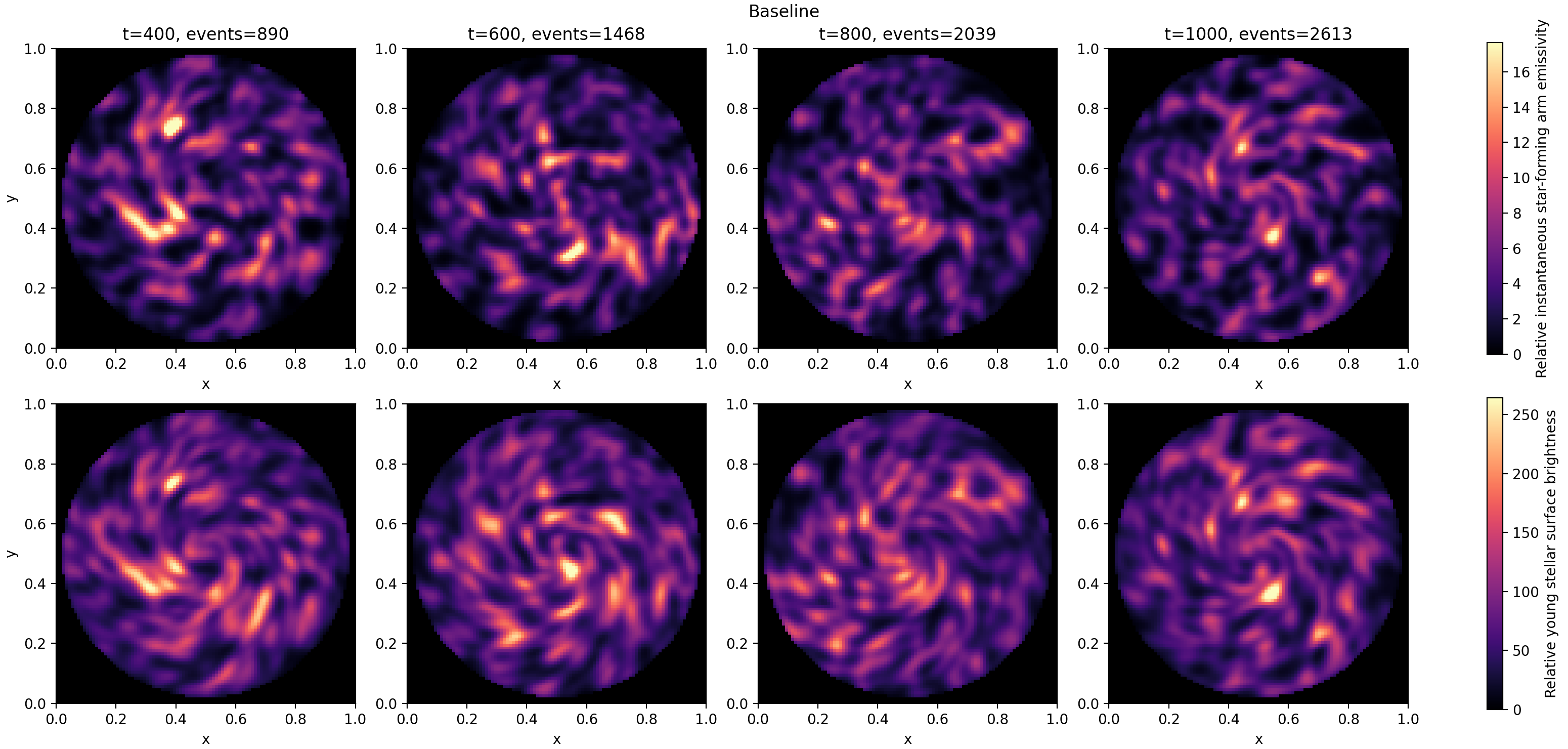}
    \caption{Representative baseline STSP--LGCHP realisation at $t=400$, $600$, $800$, and $1000$. The top row shows the relative instantaneous star-forming arm emissivity, and the bottom row shows the relative young stellar surface brightness. Transient curved and fragmented spiral-like ridges are visible in this realisation.}
    \label{fig:baseline}
\end{figure}

\begin{figure}[ht]
    \centering
    \includegraphics[width=1\linewidth]{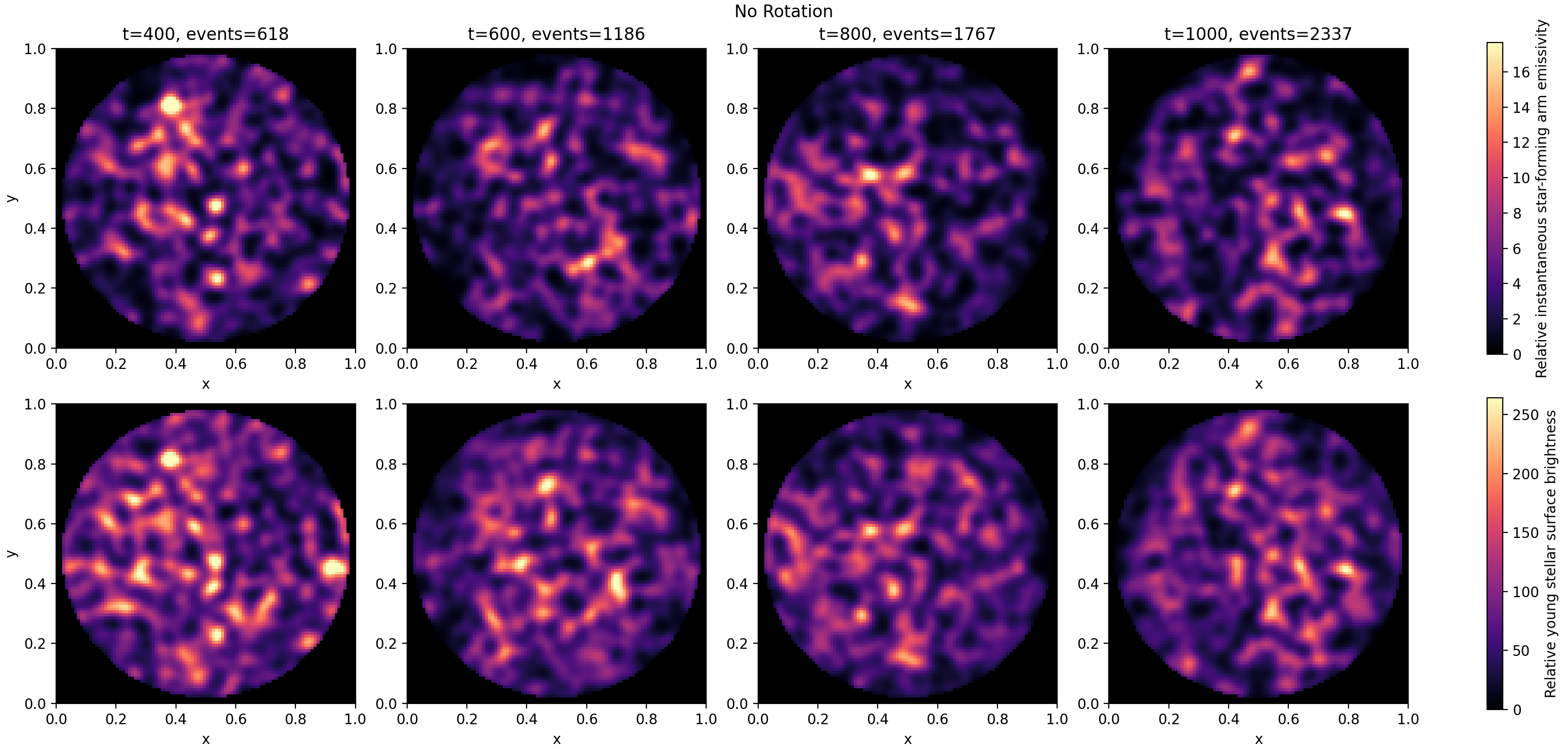}
    \caption{Representative no-rotation realisation obtained by setting $\omega_c=0$ in the event dynamics and observation-layer transport. In this realisation, the emission patterns are dominated by irregular patches and show less systematic curvature than the baseline patterns.}
    \label{fig:norotation}
\end{figure}

\begin{figure}[ht]
    \centering
    \includegraphics[width=1\linewidth]{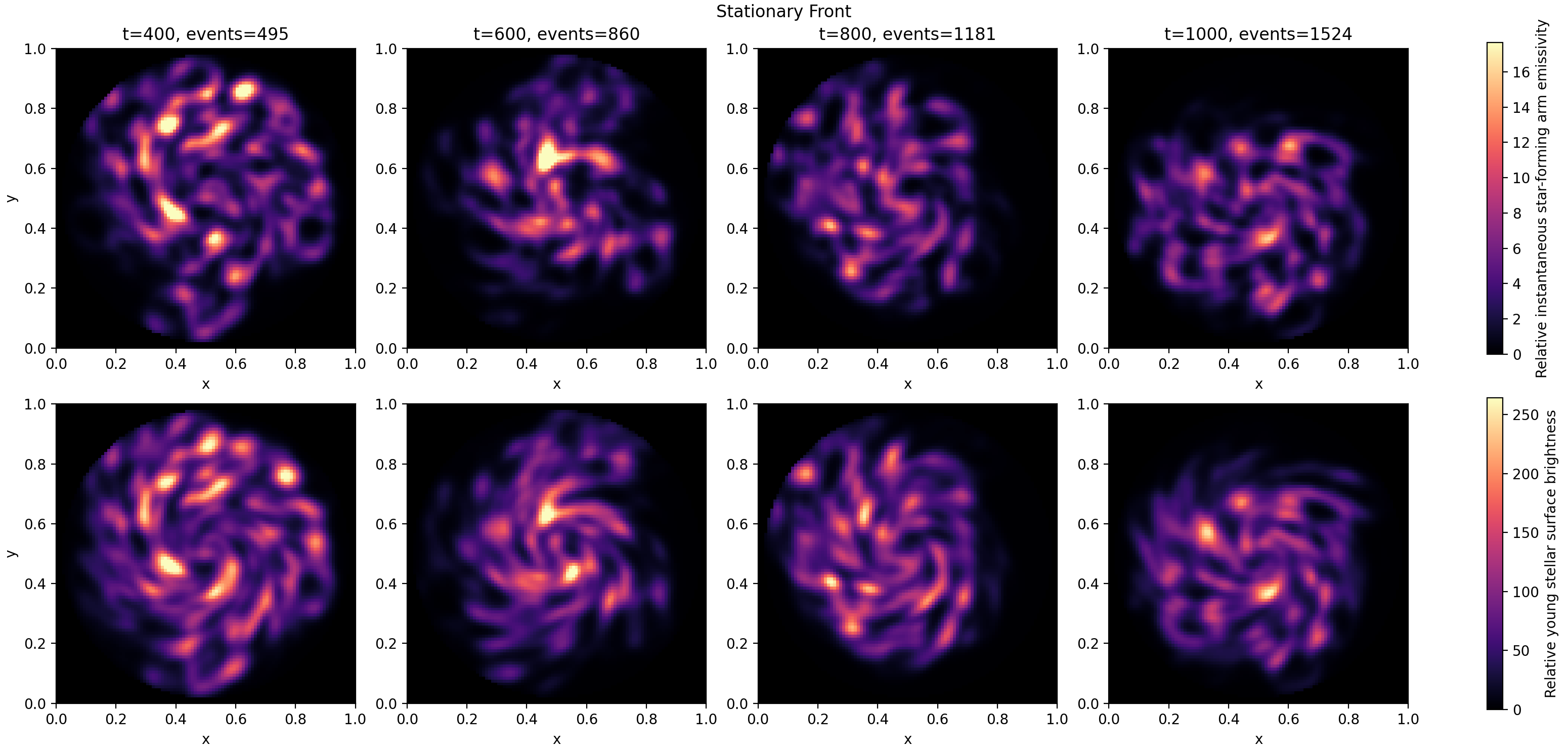}
    \caption{Representative stationary-front realisation with the front radius fixed at $v_f/\beta=0.08$. This realisation contains fewer events and shows less spatially extended emission than the baseline realisation.}
    \label{fig:stationaryfront}
\end{figure}

The cumulative event counts in Figures~\ref{fig:baseline}, \ref{fig:norotation}, and \ref{fig:stationaryfront} are $2613$, $2337$, and $1524$ at $T=1000$, respectively. Within this paired example, curved emission ridges appear in the baseline and stationary-front maps, whereas the no-rotation output is dominated by irregular patches with little coherent winding. The no-rotation case nevertheless contains a large event catalogue, suggesting that rotation influences the spatial organisation of activity more strongly than its overall abundance. This morphological comparison is based on the displayed realisations. We generate $100$ independent replicate sets, each comprising the baseline, no-rotation, and stationary-front cases, giving $300$ simulated event patterns. Within each replicate set, the three cases share the same latent-field realisation and initial history, use separate case-specific event random streams, and evolve according to their respective dynamics. For each realisation, we record the total number of generated events, $$N(T)=\sum_i\mathbf{1}\{0<t_i\leq T\},$$ and the number of recent events at each evaluation time, $$N_A(t)=\sum_i\mathbf{1}\{0<t_i\leq t,\ t-t_i\leq A\},\qquad t\in\{400,600,800,1000\}.$$ For each case, these counts are summarised by their mean, standard deviation, and $2.5\%$ and $97.5\%$ Monte Carlo quantiles. Table~\ref{tab:mc_event_counts} shows that the baseline and no-rotation cases have similar mean total counts of $2490.7$ and $2471.1$, so the larger difference between their displayed realisations is not typical of their average count behaviour. The stationary-front case has a mean total count of $1291.1$ and recent event means ranging from $91$ to $106$, compared with approximately $172$ under the baseline. Its event counts also show greater variability relative to their mean. Thus, the complete fixed-radius specification produces systematically less activity under the selected parameters. As explained above, this comparison includes the associated change in spatially integrated excitation.

\begin{table}[ht]
\centering
\begin{tabular}{|l|c|r|r|r|r|}
\hline
Case & Count & Mean & SD & $2.5\%$ & $97.5\%$ \\
\hline
Baseline & $N(T)$ & $2490.7$ & $290.9$ & $1733.4$ & $2794.0$ \\
 & $N_A(400)$ & $173.6$ & $46.1$ & $11.0$ & $266.1$ \\
 & $N_A(600)$ & $172.3$ & $25.2$ & $149.5$ & $217.7$ \\
 & $N_A(800)$ & $173.1$ & $13.5$ & $150.5$ & $199.5$ \\
 & $N_A(1000)$ & $172.4$ & $10.2$ & $151.4$ & $190.0$ \\
\hline
No rotation & $N(T)$ & $2471.1$ & $292.4$ & $1651.6$ & $2757.8$ \\
 & $N_A(400)$ & $164.6$ & $51.6$ & $8.5$ & $246.0$ \\
 & $N_A(600)$ & $172.1$ & $22.2$ & $147.0$ & $212.8$ \\
 & $N_A(800)$ & $168.9$ & $11.1$ & $149.0$ & $194.2$ \\
 & $N_A(1000)$ & $170.4$ & $10.4$ & $150.0$ & $190.6$ \\
\hline
Stationary front & $N(T)$ & $1291.1$ & $346.7$ & $280.0$ & $1729.2$ \\
 & $N_A(400)$ & $91.5$ & $60.4$ & $2.0$ & $171.1$ \\
 & $N_A(600)$ & $106.0$ & $43.8$ & $4.9$ & $168.0$ \\
 & $N_A(800)$ & $94.9$ & $29.0$ & $7.4$ & $151.0$ \\
 & $N_A(1000)$ & $101.4$ & $22.8$ & $56.9$ & $150.5$ \\
\hline
\end{tabular}
\caption{Monte Carlo summaries of the total and recent event counts based on $100$ replicates per case. The $2.5\%$ and $97.5\%$ columns give the empirical Monte Carlo quantiles.}
\label{tab:mc_event_counts}
\end{table}

\section{Discussion and Outlook}\label{sec:discussion}
This study extends the continuous point process formulation of self-propagating star formation developed by \citet{zou2025spatio} by introducing a transported excitation front, a separate local recovery field, and bounded aggregate feedback. The proposed STSP--LGCHP accommodates spontaneous formation and correlated environmental effects while representing feedback from star-forming regions through outwardly propagating excitation and local inhibition. The observation layer uses the conditional event intensity to construct continuous maps of relative instantaneous emission and recent young stellar surface brightness. No predefined spiral pattern is included in either the background model or this transformation. The repeated simulations show that removing rotation has little effect on the mean event count, while the fixed radius specification produces substantially less activity. In the displayed maps, the no-rotation case shows less coherent winding, suggesting that rotation mainly changes the spatial organisation of activity. The lower event count in the stationary-front case characterises the complete fixed radius specification, which also has lower spatially integrated excitation. The conditional intensity $\lambda_t(s)$ is the instantaneous occurrence rate of new star-forming regions per unit area and unit time. The field $E_{\mathrm{inst}}$ applies spatial smoothing to the instantaneous event rate, whereas $B_\lambda$ additionally incorporates advection and luminosity decay. Bright curved ridges in $B_\lambda$ indicate locations where recent star formation contributes strongly to the derived young stellar emission. These fields do not represent total stellar or gas density. Under the selected parameters, the displayed patterns show that the proposed mechanisms can generate flocculent arm-like features, although other processes may also contribute to spiral structure in observed galaxies.

As an initial simulation study, the present analysis uses illustrative parameters that have not been calibrated to a specific observed galaxy. The Gaussian background, axisymmetric rotation curve, isotropic latent covariance, feedback kernels, and age response are simplified representations of the underlying physical processes. The latent field and brightness integrals are evaluated on finite grids, and sensitivity to grid resolution and the initial history has not been examined. In addition, the morphology comparison is based on a single realisation and does not use a formal spiral statistic. These limitations should be addressed before drawing astrophysical conclusions from the simulated structures. Future work may develop likelihood-based or Bayesian procedures for parameter inference and neural network methods for conditional intensity estimation from observational data, while incorporating event marks, observational uncertainty, empirically measured rotation curves, and band specific observation models. Comparisons with discrete SSPSF simulations and alternative models of galactic structure could then help identify which observed patterns are characteristic of self-propagating point process dynamics. The present simulations focus on transient flocculent structures that emerge without a prescribed spiral background. However, the framework permits both deterministic and latent background components. A further extension could introduce a structured background field representing persistent large scale spiral organisation. Combining such a background with the self-propagating interaction would allow coherent grand design arms and local stochastic flocculent structure to coexist within a single model, providing a more flexible description of galaxies that exhibit both forms of spiral structure.

\section*{Acknowledgements}
The author thanks colleagues from their graduate studies at the University of Melbourne for helpful discussions related to this work.
 
\section*{Declarations}
The author declares no competing interests. No financial support was received for the preparation of this manuscript. The code and data supporting the simulation study are available from the author upon reasonable request.

\bibliographystyle{biorefs}
\bibliography{refs}
\end{document}